\documentclass[twocolumn,aps,prc,tightenlines,floats]{revtex4}

\def\bea{\begin{eqnarray}}
\def\eea{\end{eqnarray}}

\def\be{\begin{equation}}
\def\ee{\end{equation}}
\def\ba{\begin{eqnarray}}
\def\ea{\end{eqnarray}}

\usepackage{graphics}
\usepackage{graphicx}
\usepackage{epsf} 
\usepackage{slashed}
\usepackage{amsmath}
\usepackage{amssymb}

\def\sfrac#1#2{{\textstyle \frac{#1}{#2}}}

\begin{document}


\phantom{0}
\vspace{-0.2in}
\hspace{5.5in}
\parbox{1.5in}{ \leftline{JLAB-THY-08-916}}

\vspace{-1in}

\title
{\bf Nucleon and 
$\gamma N \to \Delta$ lattice
form factors in a constituent quark model}
\author{G. Ramalho$^{1,2}$ and M.T. Pe\~na$^{2,3}$ \vspace{-0.1in}  }

\affiliation{$^1$Thomas Jefferson National Accelerator Facility, Newport News, 
VA 23606 \vspace{-0.15in} } 
\affiliation{$^2$Centro de F{\'\i}sica Te\'orica de Part{\'\i}culas, 
Ave.\ Rovisco Pais, 1049-001 Lisboa, Portugal \vspace{-0.15in}}
\address{$^3$Department of Physics, Instituto Superior T\'ecnico, 
Ave.\ Rovisco Pais, 1049-001 Lisboa, Portugal }
\date{\today}

\begin{abstract}

A covariant quark model, based both on the spectator formalism and on
vector meson dominance, and
previously 
calibrated by the physical data, is here
extended to the unphysical region of the lattice data
by means of one single extra adjustable parameter -- 
the
constituent quark mass in the chiral limit.
We calculated
the Nucleon ($N$) and the $\gamma N \to \Delta$ 
form factors 
in the universe of values for that parameter described by 
quenched lattice QCD.
 A qualitative description of the Nucleon and 
$\gamma N \to \Delta$ form factors lattice data is achieved 
for light pions. 
\end{abstract}
\phantom{0}
\vspace*{0.9in}  
\maketitle


\section{Introduction}

In principle QCD is the fundamental theory 
that describes hadronic systems, but its non-perturbative 
character makes a direct application
to hadrons difficult, 
with the exception of the high energy and momentum transfer regime.
Fortunately, at low $Q^2$ effective field theories based 
on effective hadronic degrees of freedom have
been applied with success. 
Examples are chiral perturbation theory \cite{Pascalutsa06},
small scale expansion \cite{Gail06}, soliton models
and constituent quark models \cite{Pascalutsa07,NDelta,NDeltaD}.
Constituent quark models provided consistent descriptions encompassing 
the low-energy baryon spectrum, 
elastic and inelastic form factors, 
charge and magnetic radii, magnetic 
moments, axial and pseudoscalar form factors
\cite{Coester,Diaz04,Boffi02,Giannini01,Pace00,Christov96,Merten02}.

Although all those different frameworks may include the essential
features of QCD, which are confinement 
and chiral symmetry, they are only 
partial simulations of the true underlying theory.

Recently, significant progress has occurred in
calculations of QCD in the
lattice, which evaluate 
the important QCD non-perturbative 
contributions at low $Q^2$ directly from the underlying theory.
Up to now
the applications of lattice QCD 
are restricted still to large
pion masses  ($m_\pi > 350$ MeV) 
corresponding to heavy quarks,
and lattice spacings that are one order of magnitude 
smaller than the size of the nucleon ($a \sim 0.05 $ fm).
To extract information on the real world 
the results must be extrapolated both to the 
continuum limit ($a \to 0$) and 
to the physical pion mass regime \cite{Pascalutsa06,Young04}.
However, except for the 
limit to the physical region, there is 
no simple way of interpreting the lattice QCD data.

In this work we invert this procedure.
We take here the challenge of describing 
the lattice data using a quark model. 
We start with the constituent quark model obtained 
from the covariant spectator 
theory. 
Our model was presented in Refs.~\cite{NDelta,Nucleon},
where it was adjusted to the experimental data for the
four nucleon elastic form factors and   
the dominant form factor of the $\gamma N \to \Delta$ transition.
In this work we extend this model to the 
unphysical region of the lattice data.
Although the constituent quark models 
were originally thought for and applied to the 
physical limit, a constituent quark model 
that is simultaneously consistent 
with the experimental data and lattice QCD data 
is valuable, because it includes indirectly the constraints of QCD, 
and therefore satisfies
properties of the underlying theory.
A similar procedure was considered 
in Refs.~\cite{Matevosyan05a,Matevosyan05b} 
using a larger number of parameters.

Our model does not include explicit pion cloud effects.
However, as the electromagnetic 
interaction with the quarks is parametrized 
according to vector meson dominance (VMD),  
part of the pion cloud effects are indirectly taken into account.
Therefore the model can be applied in the regions
where the valence quark degrees of freedom are dominant, 
and it gives a good description of the nucleon 
and $\Delta$ systems  \cite{Nucleon,DeltaFF}.

For the nucleon, pion cloud effects are 
fundamental in the time-like region \cite{ChPT},
although they are not so important 
in the region of small positive $Q^2$ values \cite{Nucleon}, 
which is  involved in the study of
the electromagnetic transition.
For the $\gamma N \to \Delta$ transition,
as it happens also
to other constituent quark models,
without explicit pion cloud effects, the predictions for
the magnetic moment, which dominates the transition, 
deviate necessarily
from the data for low $Q^2$, 
showing that an effective pion cloud within the
VMD current 
is not sufficient. Because of the opening
of the $\pi N$ channel, an explicit
pion cloud contribution must be considered \cite{NDelta,NDeltaD}. 
Therefore our results in the physical  
low $Q^2$ region give only 
the constituent quark core 
contributions \cite{NDelta}, usually labeled
as "bare" contributions. Our results 
for these contributions are consistent with independent calculations of 
dynamical reaction models \cite{DM,Diaz07} based 
on hadronic degrees of freedom.
As for the E2 and C2 multipoles, in the large $N_c$ 
limit \cite{LargeNc} they
represent second order corrections 
and are not considered here.

As pion cloud effects in lattice QCD are expected 
to be small for $m_\pi >$ 0.40 GeV \cite{Ashley04}, 
the description of the lattice QCD data
appears as the ideal testing ground for our constituent quark model, where
pion cloud effects are not explicitly included. 
This defines the main goal of this work.
At  present, lattice calculations are
still restricted to high pion masses 
$m_\pi > $ 350 MeV, although technical improvements are 
increasingly allowing to reach nearer and nearer the physical region
\cite{Gockeler05,Alexandrou08}.
In this work we took the quenched lattice QCD data 
from Refs.~\cite{Alexandrou08,Alexandrou06}.
In the unquenched calculations the 
effects of the sea quarks are explicitly considered.
Although unquenched lattice QCD data is already available,
it is not expected that the quenched and 
unquenched data differ substantially 
in the region  $m_\pi >$ 0.40 GeV \cite{Ashley04}.
In addition, there are some differences
between unquenched data for the $\gamma N \to \Delta$ 
transition based on two different unquenched methods 
at similar pion masses  \cite{LatticeD} that 
still have to be clarified in the future.
For these reasons we took the conservative option of 
using quenched data exclusively.

Our new model presented here is based on three specific features:
i) the electromagnetic interaction 
with the constituent quarks is described within the impulse 
approximation, and considering VMD; 
ii) the wavefunctions of the quark-diquark 
system are parametrized by simple monopole 
factors reduced to the Hulthen form, with one or two 
effective range parameters that balance the details 
of the short range and the long range behavior of the system;
iii) the constituent quark magnetic anomalous moment 
scales with the inverse 
of its mass, which we write, following \cite{Cloet02}
as a function of the 
current quarks mass.

A current based on VMD is suitable to describe
the interaction of the photon with the constituent quark,
for the quark-antiquark spin-1 vertex.
Depending on the isospin, the intermediate meson pole 
corresponds to the $\rho$ or the $\omega$ meson, 
at low $Q^2$, or, in the large $Q^2$ regime,
to some other effective heavy meson pole $M_h$.
The photon interaction is then 
described as proceeding through the production of an intermediate meson state
which annihilates subsequently into a quark-antiquark pair.
VMD is successful in the description of the 
electromagnetic interaction with nucleons.

\section{Extension of the quark model to the lattice  data region}

The nucleon S-state wavefunction 
includes the correct spin structure for the quark-diquark 
spin 0 and 1 components,
associated 
to isospin 0 and 1 states, respectively. For the $\Delta$, since
total isospin is $3/2$,
the S-state wavefunction reduces to the diquark 
spin 1 with isospin 1 structure.
As for the scalar wavefunction, 
it takes the phenomenological form 
\be
\psi_B (P,k) = \frac{N_B}{m_s(\alpha_1 + \chi_B)(\alpha_2 + \chi_B)^{n_B}},
\label{eqPsiS}
\ee
where $B=N,\Delta$, $\chi_B= \sfrac{(M-m_s)^2-(P-k)^2}{Mm_s}$,
with  $M$ the baryon mass (for the nucleon or $\Delta$), 
$m_s$ is the diquark mass and $N_B$ is a normalization constant. 
[In Ref.~\cite{NDelta,Nucleon} we considered $n_B=1$ 
for the nucleon and $n_B=2$ for the $\Delta$].
The parameters $\alpha_i$ 
can be interpreted as Yukawa mass or range coefficients 
that distinguish between two different regimes for the momentum range.

The parametrization of the momentum dependence in 
terms of $\chi_B$ absorbs the 
dependence on the baryon masses $M$. 
The range parameters $\alpha_i$ for 
the nucleon and the $\Delta$ were fixed 
by the nucleon and $\gamma N \to \Delta$  
form factor experimental data
\cite{NDelta,Nucleon}.
In the spectator quark model the transition amplitude 
between a initial (momentum $P_-$) nucleon ($N$)
and a final (momentum $P_+$) baryon $B=N,\Delta$ can
be written, in impulse approximation 
\cite{NDelta,NDeltaD,Nucleon} as
\be
J^\mu =
3 \sum_\lambda 
\int_k \bar \Psi_B (P_+,k) j_I^\mu \Psi_N(P_-,k),
\ee
where $\Psi_B(P,k)$ is the complete baryon wavefunction
(including spin, isospin and momentum dependence), 
$\lambda$ is the diquark polarization 
and $j_I^\mu$ the quark current operator dependent 
of the hadron isospin $I$.
The baryon polarizations are suppressed for simplicity.
The symbol $\int_k$ represents the 
invariant integral in the on-shell diquark moment
$\int_k \equiv \int \sfrac{d^3 k}{2E_s(2\pi)^3}$, 
with $E_s$ as the diquark energy. 
The factor 3 takes account for the isospin symmetrization.

The quark current \cite{NDelta,NDeltaD,Nucleon}  
takes the general form
\ba
j_I^\mu &=&
\left( 
\frac{1}{6}f_{1+}(Q^2) + 
 \frac{1}{2}f_{1-} (Q^2)\tau_3 
\right) \gamma^\mu+ \nonumber \\
& &
\left( 
\frac{1}{6}f_{2+}(Q^2) + 
 \frac{1}{2}f_{2-} (Q^2) \tau_3 
\right) \frac{i \sigma^{\mu \nu} q_\nu}{2 M_N}. 
\ea
The constituent quark form factors $f_{i\pm}$ ($i=1,2$)
are parametrized using a VMD structure, and are normalized 
according to $f_{1\pm}(0)=1$ and 
$f_{2\pm}(0) = \kappa_\pm$, where 
$\kappa_\pm$ are defined 
in terms of the 
quark anomalous magnetic moments $\kappa_u$ and $\kappa_d$
according with 
$\kappa_+=2\kappa_u-\kappa_d$ 
and $\kappa_-=\sfrac{1}{3}(2\kappa_u+\kappa_d)$. 
See Refs.~\cite{NDelta,Nucleon} for details.

In this first calculation
the wavefunctions are reduced to S-states for the quark-diquark
system. Although it is well known that 
angular momentum components besides 
the S-states are essential for 
the description of the nucleon in Deep Inelastic Scattering, 
their effects are not so evident 
in the elastic form factors, particularly for low $Q^2$ \cite{Nucleon}.
As for the $\Delta$, calculations of the valence quark 
contribution associated with higher angular 
momentum states (D-states) suggest a small effect \cite{NDeltaD,LatticeD}.
In fact, even with only S-wave 
components in the wavefunctions for both the nucleon
and the $\Delta$, the model generates the
dominant contribution for the $\gamma N \to \Delta$ transition, 
the dipole magnetic moment form factor.
The non-zero contributions for the
other form factors appear only when D states are
included, or when pion cloud effects are taken in consideration
\cite{NDeltaD,LatticeD}.
At the end, we will estimate the effect in the 
$\gamma N \to \Delta$ quadrupole
transition form factors, in the lattice regime, 
from adding the D-states to the present model.

As shown in \cite{Nucleon} the diquark mass scales out 
from the formulas obtained for the electromagnetic form factors.
This allowed us to ignore any explicit 
dependence on the quark mass. 
This mass dependence was present
in the nucleon, $\Delta$ and vector meson masses in an implicit way
only.
However, in order to grant a comparison 
of our results with the lattice data,
we
extend here the model to include a dependence of
the quark anomalous moment $\kappa_u$ and $\kappa_d$ on the quark mass. 
This dependence was not explicitly considered  
in Refs.~\cite{NDelta,Nucleon} because they dealt 
only with the physical data.


\begin{figure*}[t]
\vspace{.5cm}
\centerline{
\mbox{
\includegraphics[width=7.05cm]{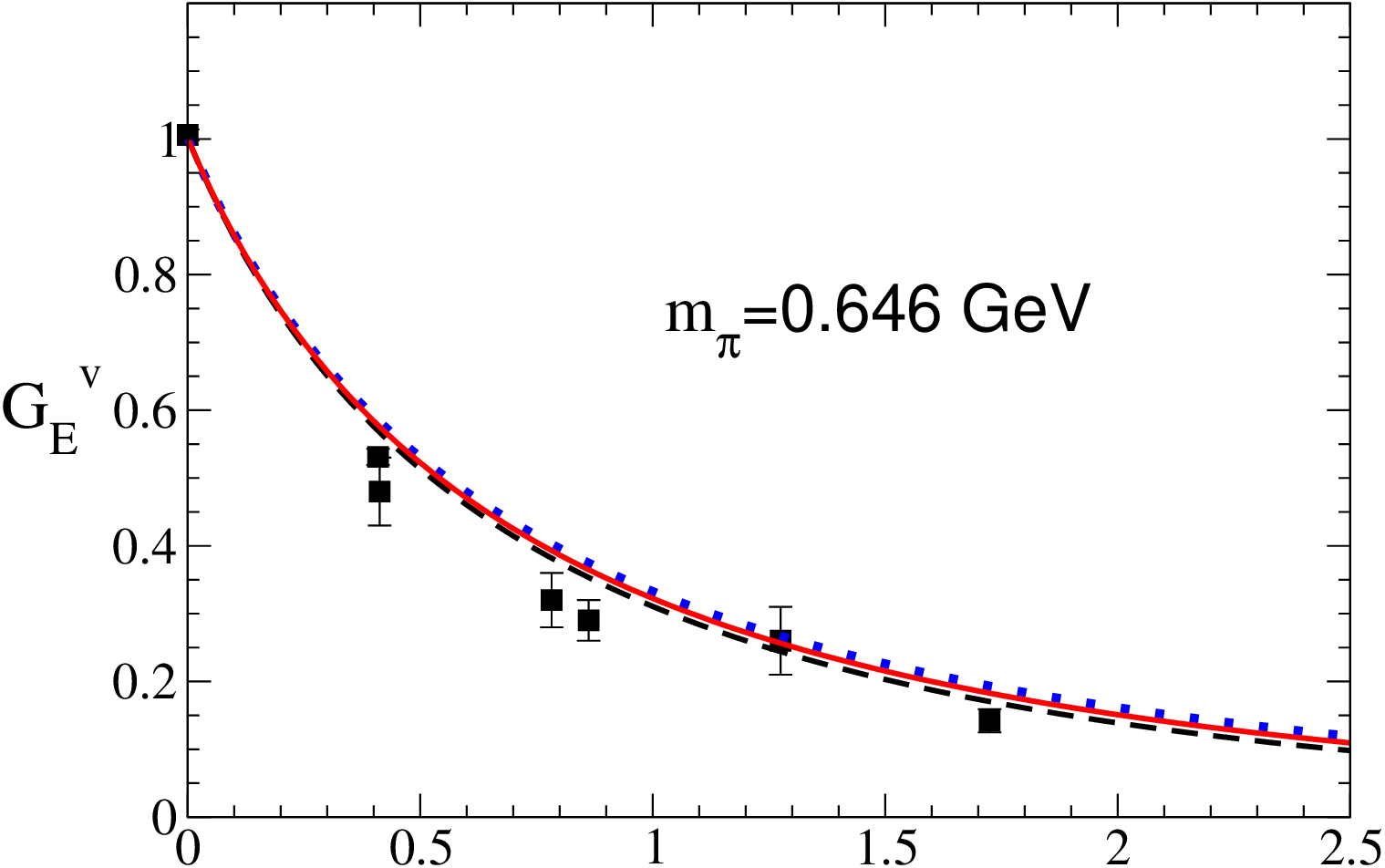} \hspace{.1cm}
\includegraphics[width=7.05cm]{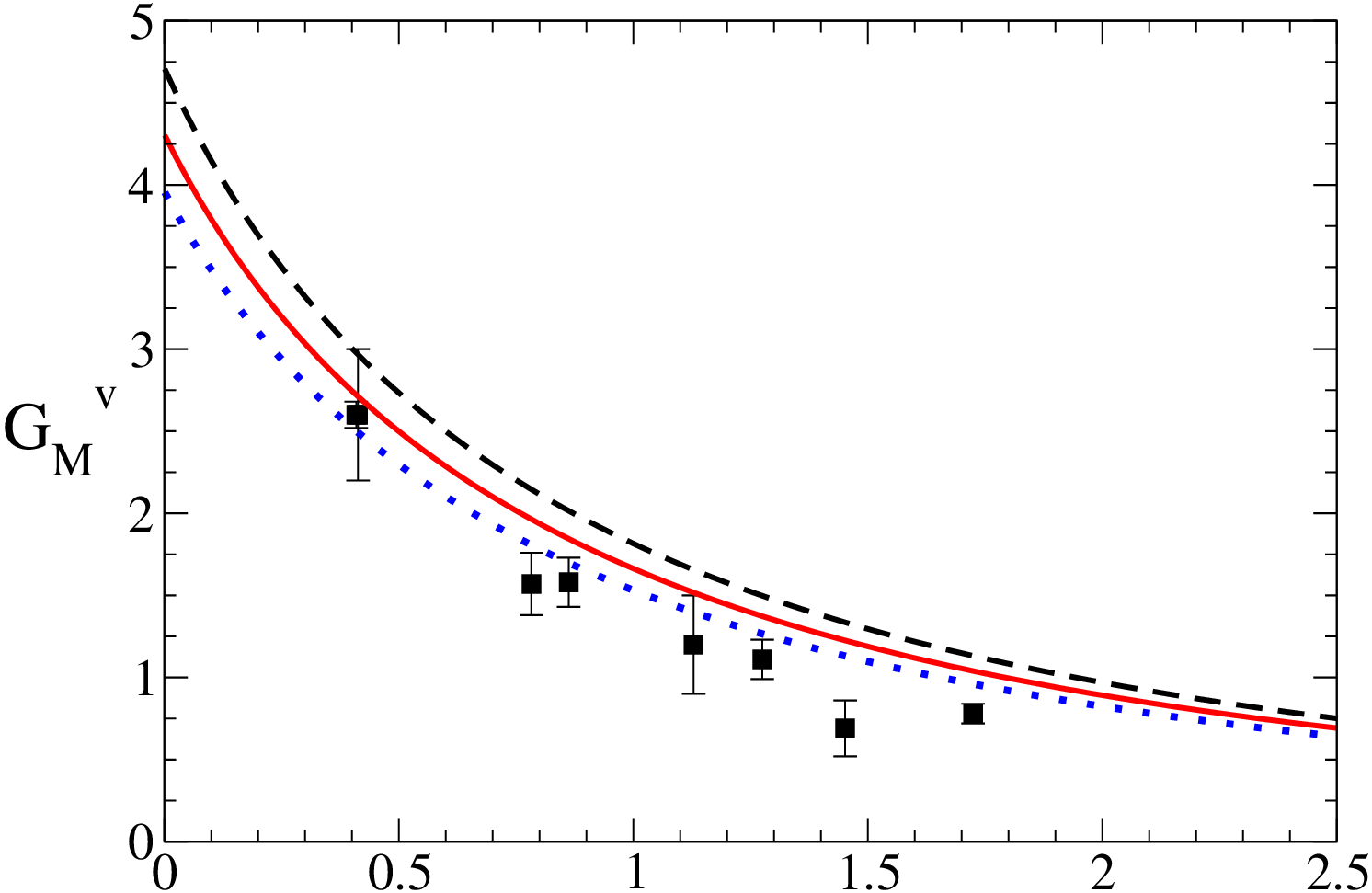}} }
\vspace{.8cm}
\centerline{
\mbox{
\includegraphics[width=7.05cm]{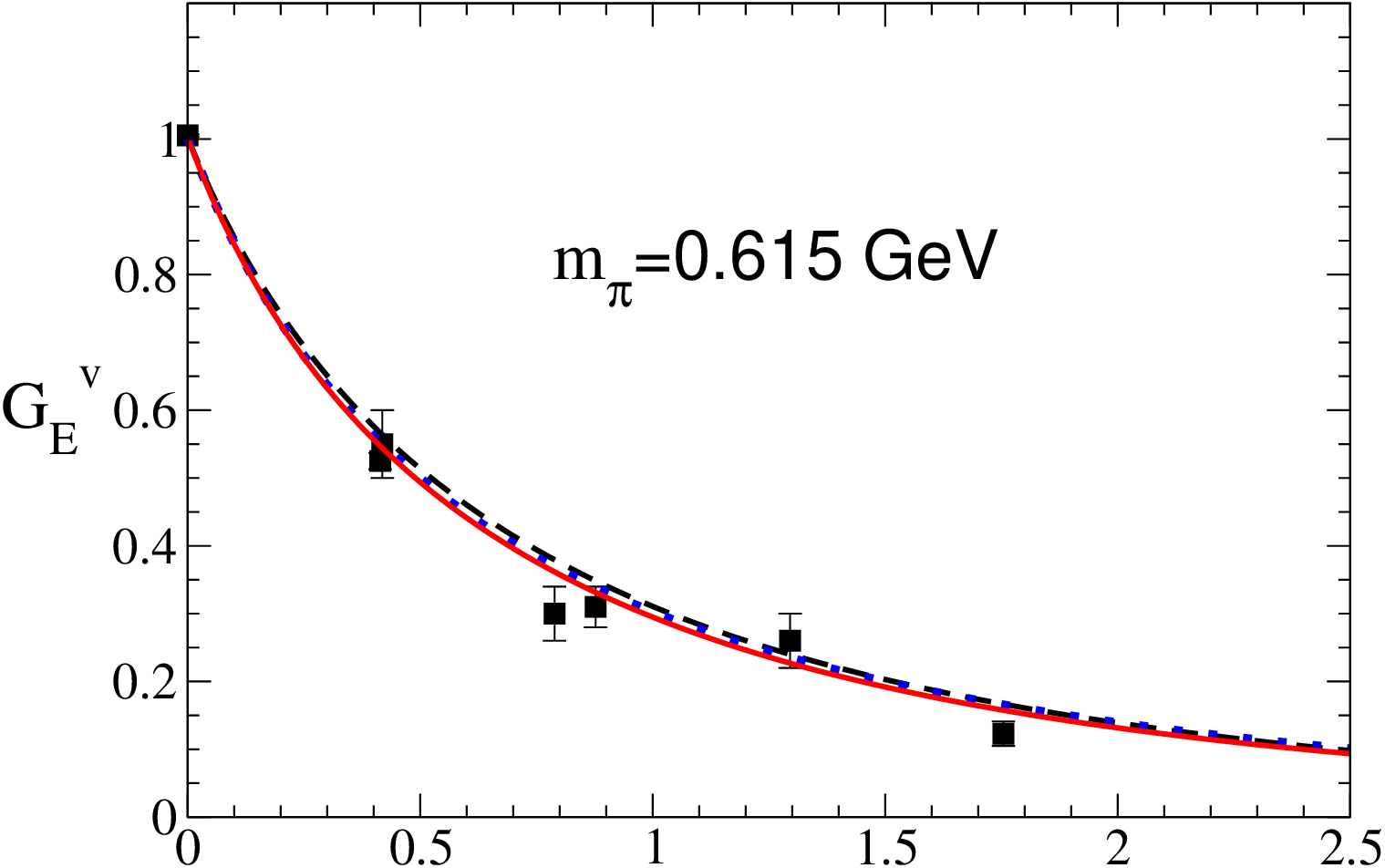} \hspace{.1cm}
\includegraphics[width=7.05cm]{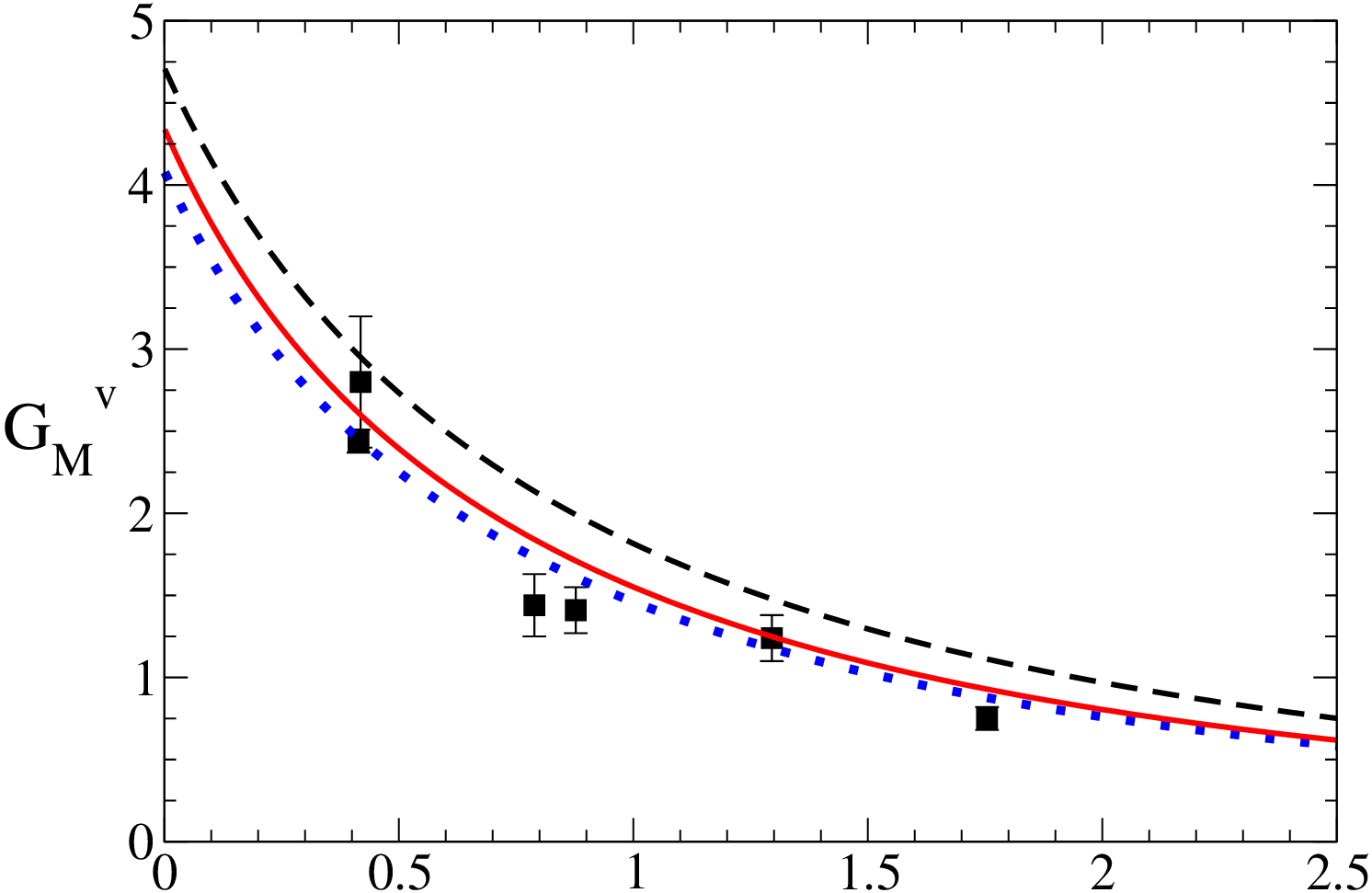}} }
\vspace{.8cm}
\centerline{
\mbox{
\includegraphics[width=7.05cm]{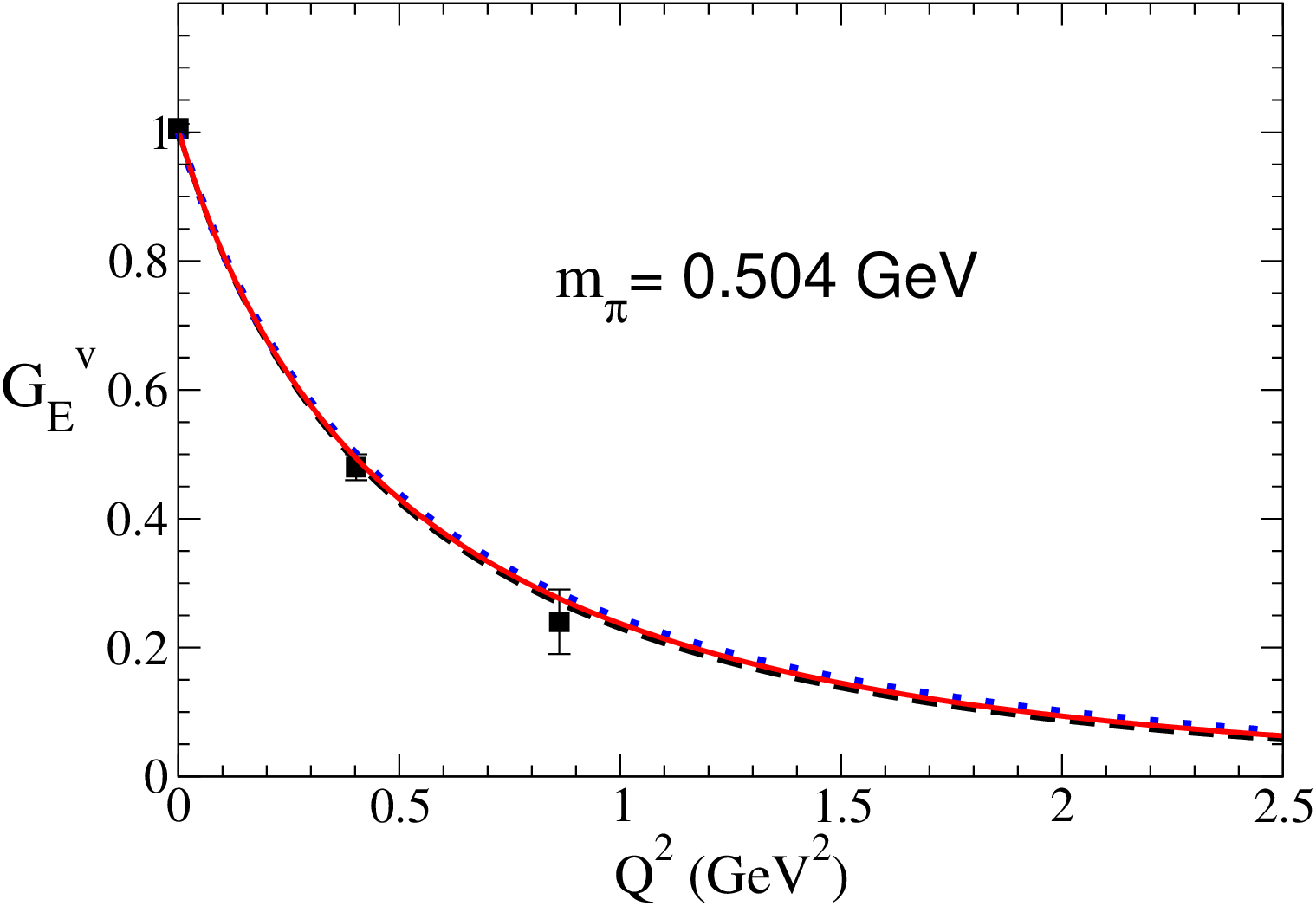} \hspace{.1cm}
\includegraphics[width=7.05cm]{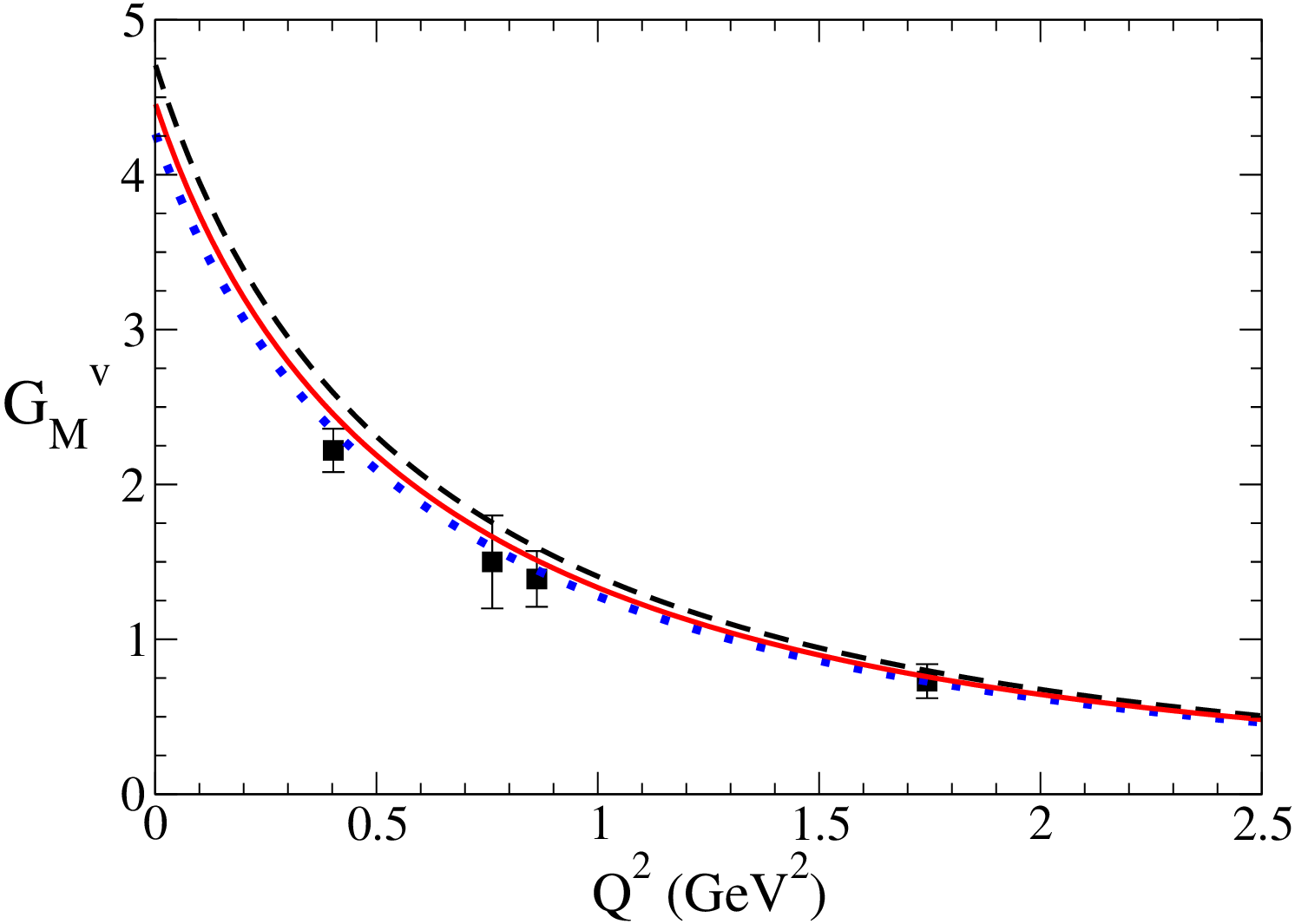}} }
\vspace{-.2cm}
\caption{
Nucleon isovector form factors in lattice QCD from Ref.~\cite{Gockeler05}
and our predictions corresponding respectively to $M_\chi$
$+ \infty$ (dashed line), 0.42 GeV (dotted line) and 0.80 GeV (solid line).}
\label{figNucleonFF}       
\end{figure*}

Inspired by Ref.~\cite{Cloet02}, which combines
chiral symmetry with conventional quark models, and applies
an analytic continuation of the chiral expansion to the simple
SU(6) model, we then use the 
smooth variation of the hadronic properties with the current quark masses
above 60 MeV. Therefore,
in the spirit of the constituent quark models,
we assume here that $\kappa_q$  ($q=u,d$) scales with $1/M_q$,
where $M_q$ is the constituent quark mass.
Labeling the quark anomalous moment
at the physical point by $\kappa_q^0$, we can write $\kappa_q$, for 
an arbitrary constituent quark mass $M_q$, as 
\be
\kappa_q= \frac{M_q^{\rm phy}}{M_q} \kappa_q^0,
\label{eqKq}
\ee
where $M_q^{\rm phy}$ is the constituent quark 
mass at the physical point.

To include an explicit dependence on the quark mass we 
consider the parametrization due to Cloet {\it et.~al.}~\cite{Cloet02}

\be
M_q \,=\, M_\chi+ cm_q \,=\, M_\chi + c m_q^{\rm phy} 
\frac{m_\pi^2}{(m_\pi^{\rm phy})^2},
\label{eqMq}
\ee 
where $m_q$ ($m_q^{\rm phy}$) is the (physical) 
current quark mass, $M_\chi$ 
is a new parameter corresponding to the 
constituent quark mass in the chiral limit ($m_q=0$),
and $c$ is a coefficient of the order of the unity.
In  the same equation $m_\pi$ stands for 
the pion mass in the model, a parameter to be fixed by the lattice data,
and $m_\pi^{\rm phy}$ for the physical mass.
Following  Ref.~\cite{Cloet02} we considered
$c m_q^{\rm phy} = 5.9$ MeV.

The parametrization in (\ref{eqMq}) is most sensitive to $M_\chi$.  
Reference \cite{Cloet02} fixes $M_\chi=0.42$ GeV.
Different descriptions using quark models and  
different lattice sizes lead to different values \cite{Matevosyan05a}.
For this reason we use $M_\chi$ as the only free parameter allowed to vary
in the calculation presented here.
It is needed to introduce an explicit dependence 
of the nucleon magnetic moment on the pion mass, 
and to enable the connection of the constituent quark model 
to the lattice QCD calculations 
\cite{Young04,Gockeler05,Alexandrou08,Alexandrou06,Leinweber01,Wang07}.

We consider three different cases.
First we considered the case $M_\chi=+\infty$,
corresponding to the limit where $\kappa_q$ has no dependence on $m_\pi$.
We tested also the original parametrization
$M_\chi=$ 0.42 GeV \cite{Cloet02}.
Finally, although we did not 
performed a systematic fit, we tested 
several other values of  $M_\chi$.

Furthermore, to extend the spectator model to the region of the
quenched lattice QCD data we still need to consider 
the nucleon and $\Delta$ masses determined by 
the lattice simulations \cite{Alexandrou08}. This brings to the calculation
an implicit dependence on the pion mass. 
As for the vectorial mesons included in the VMD, 
quark current picture, we use the parametrization \cite{Leinweber01}
\be
m_\rho=
c_0 + c_1 m_\pi^2,
\label{eqMrho}
\ee
where $c_0=0.776$ GeV and $c_1= 0.427$ GeV$^{-1}$.
The simple parametrization (\ref{eqMrho}) 
describes well the quenched lattice QCD data
and is consistent with finite 
volume corrections \cite{Young04,Allton06}.
In this applications we consider 
this parametrization together with 
model II of the Refs.~\cite{NDelta,Nucleon}, 
where the heavy vectorial meson mass is $M_h= 2 M_N$,
with $M_N$ the nucleon mass of the lattice calculations.

\section{Results}

In Fig.~\ref{figNucleonFF} we compare the predictions 
of our model {for the nucleon form factor with the
lattice data} from Ref.\ \cite{Gockeler05},
corresponding to the three values of $M_\chi$.
We consider in particular the isovector  
nucleon form factor because the contributions of the 
disconnected diagrams vanishes in lattice calculations, if the 
flavor SU(2) symmetry is assumed as in Ref.~\cite{Gockeler05}.
We considered the lowest pion masses and the smallest 
lattice spacings.
In particular, we take the data corresponding to
$m_\pi=$ 504 MeV, 649 MeV ($a=0.059$ fm) and 
$m_\pi= 615$ MeV  ($a=0.078$ fm).
For larger lattice spacings a dependence on $a$ is observed 
\cite{Gockeler05}.
From the figure we conclude that 
$M_\chi=0.42$ GeV and 0.80 GeV gives an excellent description 
of the nucleon data.

Still, when the pion mass increases there 
is a systematic deviation of our model 
from the lattice data of Ref.~\cite{Gockeler05}.
This deviation may be a consequence of the fact that
the wavefunction parametrization 
in terms of a low momentum scale ($\alpha_1^N$) (long range behavior) 
and high momentum scale ($\alpha_2^N$) was kept unchanged.
As the pion masses vary, at least the 
long range parameter may vary, 
and may vary more rapidly
for larger pion masses. 
As for the short range parameter ($\alpha_2^N$)
it sets the scale below which  
the short range physics becomes important.
As observed in applications of the finite-range regularization 
effective field theory, the scale 
associated to the momentum cut-off 
(and consequently short range effects) can be expressed by 
an universal regulator which describes 
simultaneously the large pion mass 
regime and the physical regime 
\cite{Young04,Wang07,Thomas04}. In our calculation
 $\alpha_2^N$ 
is related to that universal regulator, and it is
expected not to depend crucially on the pion mass value.

\begin{figure}[t]
\vspace{.7cm}
\centerline{
\mbox{
\includegraphics[width=9.0cm]{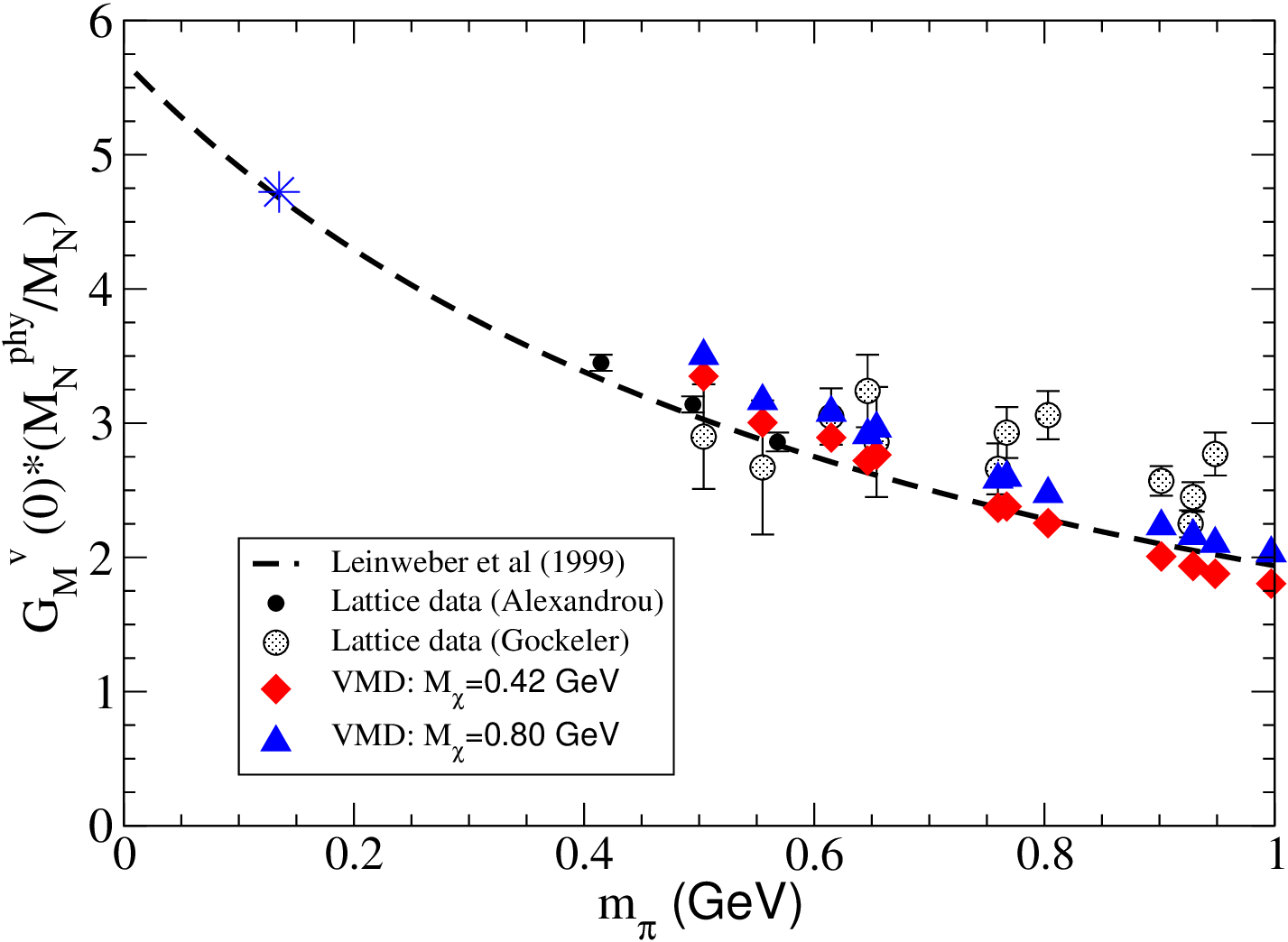} }}
\vspace{-.1cm}
\caption{Nucleon
isovector magnetic moment as function 
of $m_\pi$, in physical nucleon magnetons.
Theoretical line from Ref.~\cite{Leinweber99}. Our results are labeled "VMD" and shown for
two values of the $M_\chi$ parameter.}
\label{figGMv}
\end{figure}

For a finer analysis of our results,
we compare our predictions for 
the isovector nucleon magnetic moment 
in physical nucleon magnetons, 
with the lattice QCD data and the 
chiral result from Ref.~\cite{Leinweber99}.
To convert $G_M^v(0)$ given by the 
lattice data in units $\sfrac{e}{2M_N}$
($M_N$ nucleon mass in lattice)  
to "physical" units $\sfrac{ e}{ 2M_N^{\rm phy}}$,
we need to use the transformation
$G_M^v(0) \to \sfrac{ M_N^{\rm phy}}{ M_N}G_M^v(0)$. 
The results are presented in the Fig.~\ref{figGMv}
for $M_\chi=0.42$ GeV and   $M_\chi=0.80$ GeV, as function of $m_\pi$.
The agreement of our results with the 
chiral expression from Ref.~\cite{Leinweber99}  
shows the consistency of our calculations with 
chiral calculations.

\begin{figure*}[t]
\vspace{.7cm}
\centerline{
\mbox{
\includegraphics[width=7.3cm]{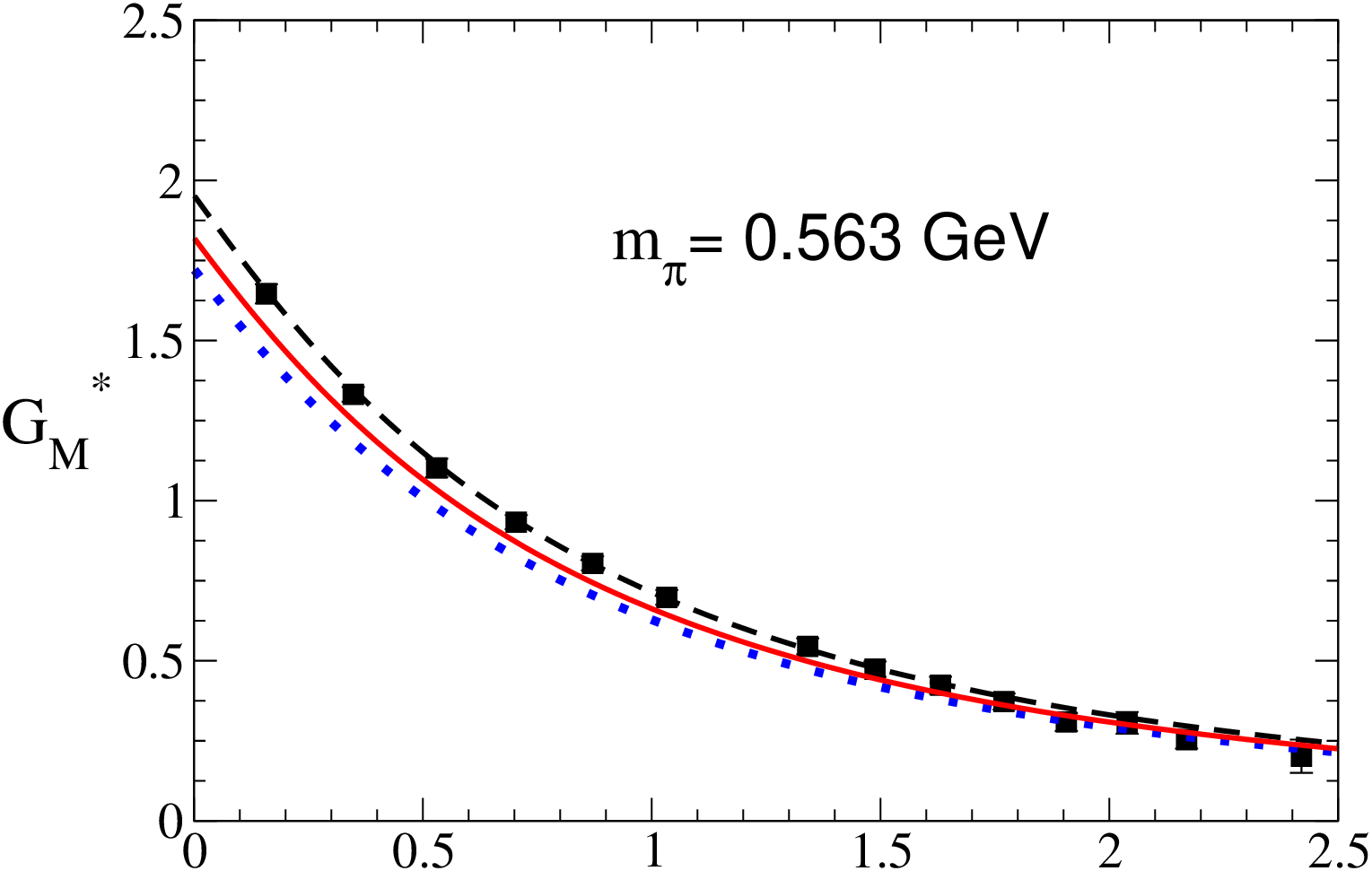} \hspace{.05cm}
\includegraphics[width=7.3cm]{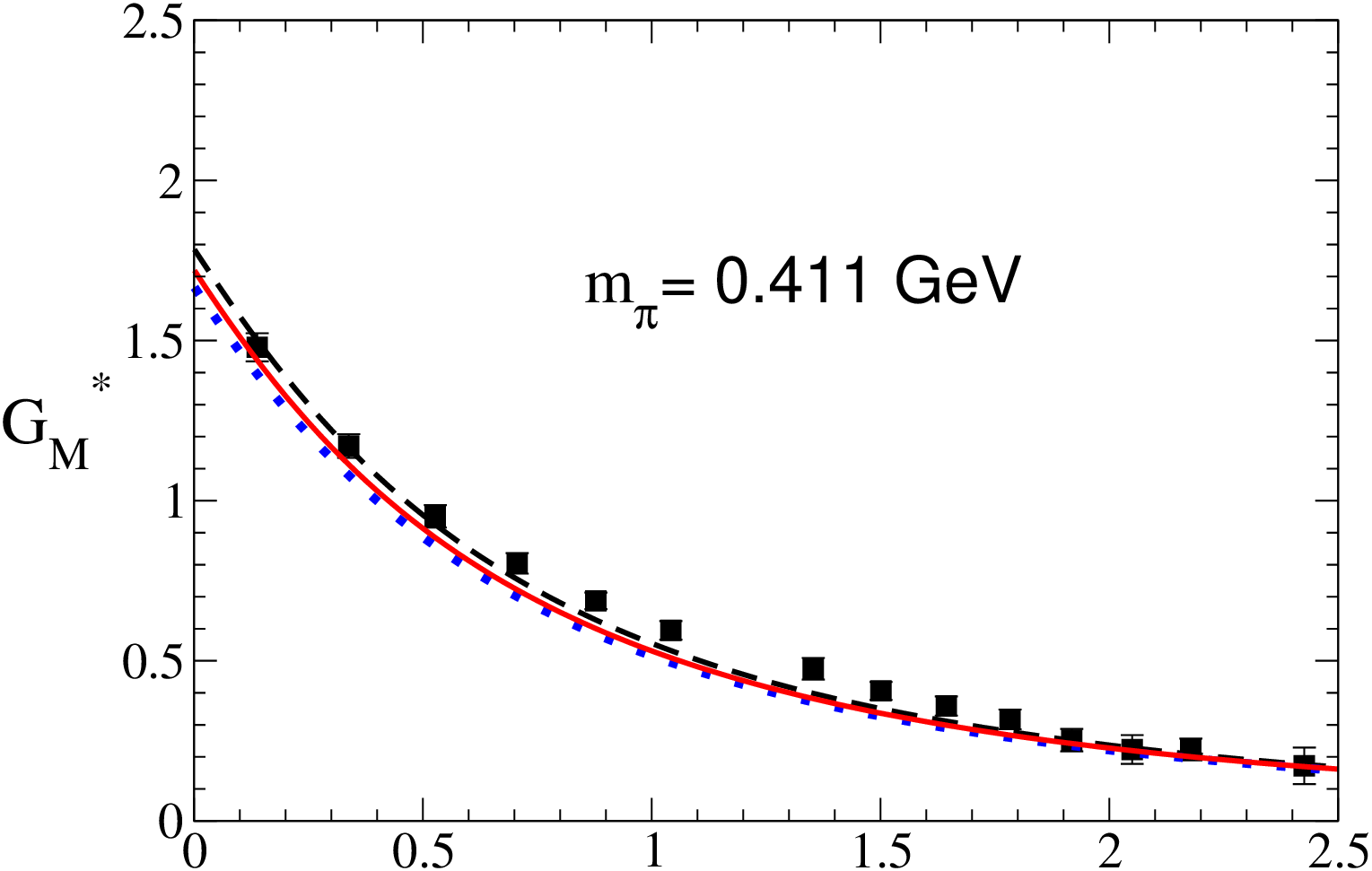}} }
\vspace{.8cm}
\centerline{
\mbox{
\includegraphics[width=7.3cm]{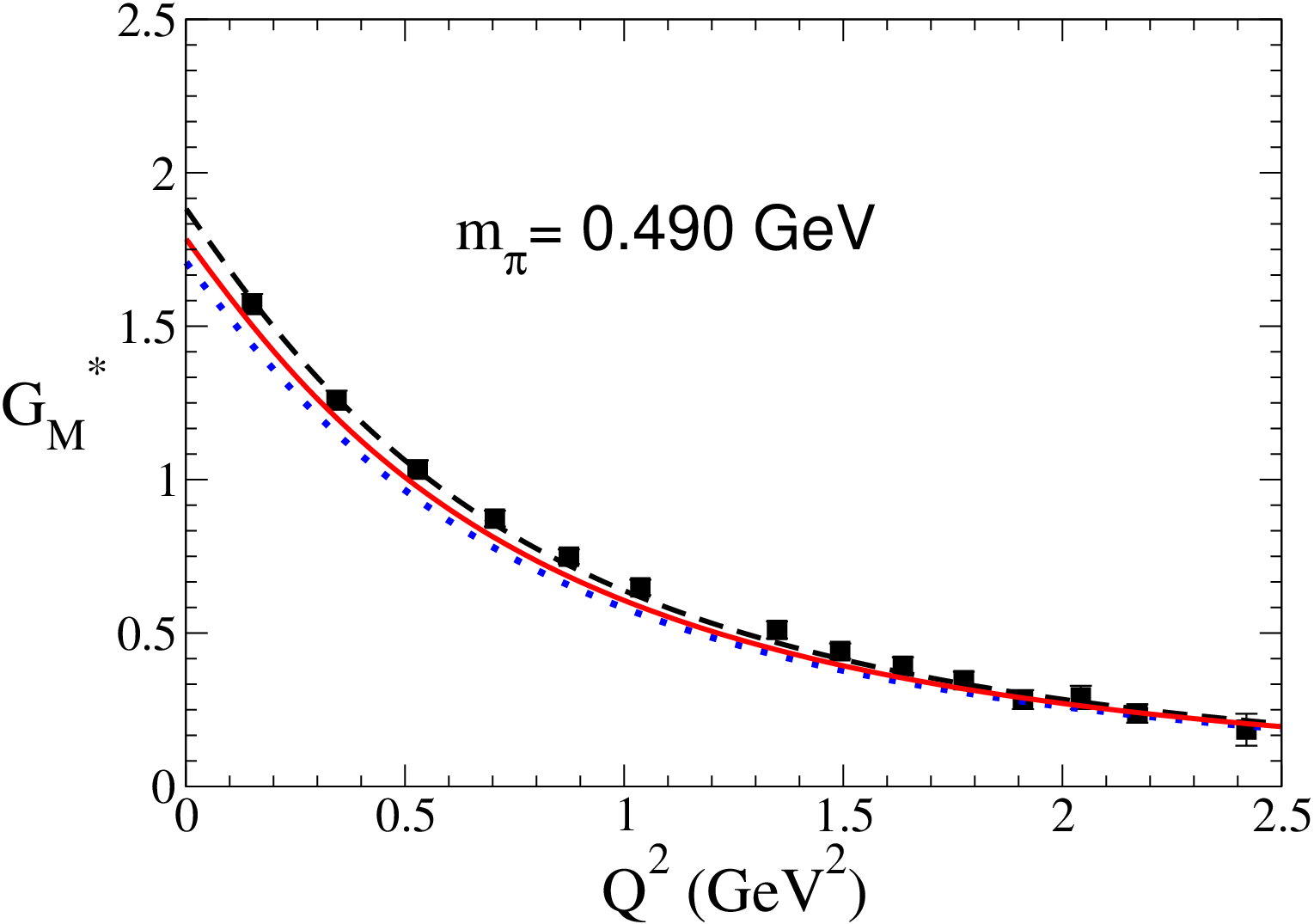} \hspace{.05cm}
\includegraphics[width=7.3cm]{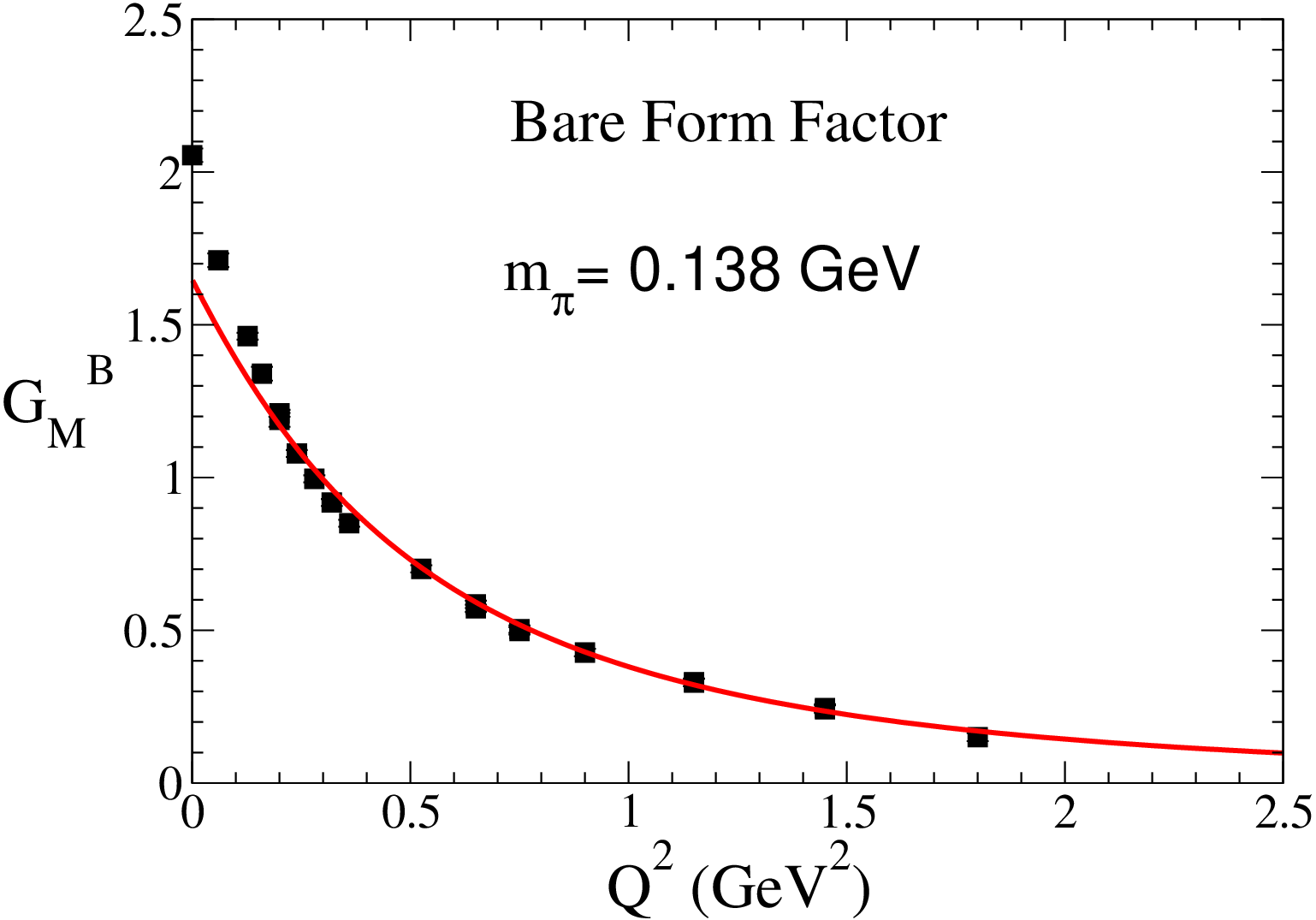}} }
\vspace{-.1cm}
\caption{
$\gamma N \to \Delta$ magnetic form factor 
lattice data from Ref.\ \cite{Alexandrou08} 
and 'bare' form factors from Ref.~\cite{Diaz07} (physical point) 
compared with models $M_\chi= 0.42$ GeV, 0.80 GeV  
and $M_\chi=+ \infty$.
The conventions are the same as in Fig.~\ref{figNucleonFF}. }
\label{figNDeltaFF}
\end{figure*}

Finally, in Fig.~\ref{figNDeltaFF}
we compare the quenched lattice data for the 
$\gamma N \to \Delta$ transition from Ref.~\cite{Alexandrou08}
with the spectator quark model 
corresponding to $M_\chi= 0.42$ GeV, 0.80 GeV and $M_\chi= + \infty$. 
All cases shown correspond to a lattice spacing of $a=0.092$ fm.
The contribution of the quark core extracted 
from \cite{Diaz07} at the physical point is also included. 
To be consistent, we considered the 
parametrization of Eq.~(\ref{eqMrho}) for $m_\rho$,  
although the original Ref.~\cite{Alexandrou08} gives slightly  
different results.
As for $M_N$ and $M_\Delta$, we use the 
values derived directly from the lattice data
\cite{Alexandrou08}.
For the larger pion masses we observe 
an almost perfect agreement between 
the predictions of $M_\chi = + \infty$ and the data,
although $M_\chi= 0.80$ GeV is also close.
For $m_\pi=0.411$ GeV we have also a good agreement,
except for a slight deviation from the lattice data 
for $0.7$ GeV$^2 < Q^2 < $ 2 GeV$^2$.
This deviation may result from the  
effect of the pion cloud for light pions,
predicted to be important for $m_\pi < 0.40$ GeV 
\cite{Ashley04}. 
As for the physical pion mass case, our model 
is coherent
with the constituent quark core data labeled 'Bare Form Factor', 
which is extracted indirectly from experiment \cite{NDelta,Diaz07}.
In this case all lines coincide.

As mentioned already, the D-states 
in the $\Delta$ wavefunction 
induce contributions for the subleading electric and Coulomb 
quadrupole transition 
form factors $G_E^\ast$ and $G_C^\ast$.
The exact contributions depend on the 
specific parametrization, in particular on the 
admixture coefficients for the two D-waves.
In Ref.~\cite{LatticeD} it was shown how
a percentage of D-states of $\approx 0.9\%$
can provide an  excellent description of the quadrupole 
lattice data, without significantly changing the description of the
dominant magnetic dipole form factor. 
That application corresponds to the $M_\chi=+ \infty $ limit 
and estimates the effect of the valence quark 
(or bare) contributions as only $\approx 20\%$ 
of the total quadrupole for both $G_E^\ast$ 
and $G_C^\ast$, at the physical point
(the remaining contribution being the pion cloud).
On the other hand, in that particular parametrization, and
in the region of study $Q^2<1.5$ GeV$^2$,
the effect of the constituent quark mass dependence
described by Eq.~(\ref{eqKq}) corresponds 
to a correction of less than 20\% 
for both form factors, a correction smaller than the typical 
lattice errorbands ($\approx 30\%$).

\section{Conclusions}

Chiral perturbation methods 
for lattice QCD extrapolations 
are useful for small $Q^2$ (like $Q^2 < 1.5$ GeV$^2$),  
but are not adequate for the high $Q^2$ region.
Constituent quark models can supply an alternative guidance
for lattice QCD extrapolations.
In this work we consider a covariant constituent quark model
of the nucleon and the $\Delta$ based on the spectator formalism
for the quark-diquark system
(covariance is an important issue
for the description of 
the high momentum transfer processes).

The model contains no explicit pion cloud, 
besides the effects included in the $\rho$ term of the VDM picture
for the electromagnetic current, 
and is therefore reduced to the bare quark hadron
structure.
Since pion cloud effects are expected to be negligible for large values of the
pion mass,  
the comparison of the model
to the quenched data for pion masses larger than 450 MeV is justified a priori.
To accomplish that comparison
the initial covariant quark model was extended 
to the lattice data region, $m_\pi < 700$ MeV.
In light of the work in Ref. \cite{Cloet02}, 
this extension was done by introducing the constituent quark mass in 
the chiral limit parameter, $M_\chi$, and the nucleon $\rho$, $\Delta$ masses
used in the lattice calculations.

The main conclusion is that
the covariant constituent quark model which was previously calibrated
by means of a quantitative description of the 
nucleon, $\gamma N \to \Delta$ and $\Delta$ form factor data
in the physical region, 
as shown in previous works \cite{NDelta,Nucleon,DeltaFF},
after a simple extension involving one parameter only, 
describes also quantitatively well
the lattice results for the nucleon isovector form factor
and the $\gamma N \to \Delta$ "bare" magnetic form factor.
With $M_\chi=0.42$ GeV, 
consistently
with the range $M_\chi =0.3-0.45$ GeV suggested
by several constituent quark models
\cite{Coester,Diaz04,Giannini01,Christov96}, we
obtain a very good description of the lattice
nucleon form factors data, but underestimate
the $\gamma N \to \Delta$ magnetic moment form factor data
by less than 9\% at low $Q^2$ for $m_\pi < 500$ MeV.
Note that the original parametrization $M_\chi=0.42$ GeV
\cite{Cloet02} was a result of a phenomenological 
fit and was not derived from first principles.

An optimal description of the lattice data 
for both nucleon and  $\gamma N \to \Delta$ transition 
form factors is achieved,
once the scale of the constituent quark mass in the chiral limit is fixed
as $M_\chi=0.80$ GeV.
[The values $M_\chi=0.42$ and 0.80 GeV 
are better for the nucleon data; 
$M_\chi=0.80$ GeV and $+\infty$ are better for 
the $\gamma N \to \Delta$ data].
The parameter $M_\chi=0.80$ GeV is relatively large when compared
with alternative constituent quark models,
which may indicate that some of the
effective quark-antiquark configurations contained 
in our model through the VMD 
mechanism in the electromagnetic current,
for instance, do not correspond to the 
lattice calculations in the quenched approximation.
Still, the possibility of adjusting other parameters is very promising.



A refinement of the description
can be pursued, by increasing the number of adjustable parameters, 
or simply by introducing an extra dependence on the pion mass.
For instance, in the future we certainly plan to check the 
assumption that the $\alpha_1$ parameter which controls
the long range regime does not depend on the pion mass value.
In addition, we may consider as well an explicit dependence  of the 
heavier meson mass pole $M_h$, that regulates the short range effect 
in VMD mechanism, on the pion mass.
Alternatively, a pion mass dependence of the $\alpha_2$ parameter can 
be introduced.
Furthermore, we want to study the 
quality of the description
in the high pion masses region, not probed yet here.


\vspace{0.2cm}
\noindent
{\bf Acknowledgments:} 

The authors are particularly thankful 
to G\"{o}ckeler for supply information about Ref.~\cite{Gockeler05}.
G.~R.\ would like to thank specially  
to Ross Young for the detailed explanations 
of the lattice proprieties and the extrapolations 
for the real world. 
G.~R. also thank Franz Gross, Ian Cloet, Michael Pardon,  
Anthony Thomas and Ping Wang for helpful discussions.
This work was partially supported by Jefferson Science Associates, 
LLC under U.S. DOE Contract No. DE-AC05-06OR23177.
G.~R.\ was supported by the portuguese Funda\c{c}\~ao para 
a Ci\^encia e Tecnologia (FCT) under Grant No.~SFRH/BPD/26886/2006. 
This work has been supported in part by the European Union
(HadronPhysics2 Project ``Study of Strongly Interacting Matter'').



\end{document}